\documentclass[doublecol]{epl2}
\usepackage{graphicx}
\usepackage{bm}
\usepackage{amssymb}
\usepackage{amsmath}
\usepackage{braket}
\usepackage[colorlinks=true,linkcolor=blue,citecolor=blue,urlcolor=blue]{hyperref}

\definecolor{mymagenta}{rgb}{1.0,0.0,1.0}
\definecolor{mycyan}{rgb}{0.0,1.0,1.0}
\definecolor{myyellow}{rgb}{1.0,1.0,0.0}
\definecolor{myorange}{rgb}{1.0,0.27,0.0}

\title{Light cone in the two-dimensional transverse-field Ising model\\
in time-dependent mean-field theory}
\shorttitle{Light cone in the 2D-TFIM\dots} 

\author{J. Hafner\inst{1}\thanks{E-mail: \email{jonas@lusi.uni-sb.de}}, B. Bla{\ss}\inst{1}\thanks{E-mail: \email{bebla@lusi.uni-sb.de}} \and H. Rieger\inst{1}\thanks{E-mail: \email{h.rieger@mx.uni-saarland.de}}}

\shortauthor{J. Hafner \etal}

\institute{\inst{1} Theoretical Physics, Saarland University, 66123 Saarbr{\"u}cken, Germany}

\date{\today}

\abstract{We investigate the propagation of a local perturbation in the two-dimensional transverse-field Ising model with a time-dependent application of mean-field theory based on the BBGKY hierarchy. We show that the perturbation propagates through the system with a finite velocity and that there is transition from Manhattan to Euclidian metric, resulting in a light cone with an almost circular shape at sufficiently large distances. The propagation velocity of the perturbation defining the front of the light cone is discussed with respect to the parameters of the Hamiltonian and compared to exact results for the transverse-field Ising model in one dimension.}


\begin{document}
\maketitle

\section{Introduction}
The relaxation process of isolated many-body quantum systems has gained tremendous interest in recent years (see for example Refs.\,\cite{Polkovnikov2011,Eisert2015,Essler2016} for an overview). Here one especially tries to answer the following two questions: (I) How does a perturbation propagate through a system driven out of equilibrium? (II) Does the system evolve towards a stationary state and, if yes, what is its nature? With respect to the first question one is interested in the propagation velocity of a local perturbation. Lieb and Robinson have shown that for nonrelativistic many-body quantum systems described by translation-invariant Hamiltonians with only finite-range interaction terms there is an upper bound of the velocity, the so-called Lieb-Robinson bound (LRB), which depends only on the parameters of the Hamiltonian and is independent of the wave function of the system \cite{Lieb1972}. Thus even for highly entangled states with long-range interactions a local perturbation needs a finite time to reach distant points of the system and the existence of a light cone emerges. While in one-dimensional systems the light cone is fully determined by the propagation velocity of the perturbation, in higher-dimensional systems its geometry has to be studied as well, i.e. one has to answer the question to which metric the propagation of the perturbation through the system obeys. While for continuum models one would expect the Euclidian metric
\begin{align}
d_{\text{Eucl}}(\mathbf{r})=\sqrt{\sum_{i=1}^{D}r_{i}^{2}}\;,
\end{align}
for systems on orthogonal lattices with interactions only between nearest neighbours one may think of the Manhattan metric
\begin{align}
d_{\text{Manh}}(\mathbf{r})=\sum_{i=1}^{D}|r_{i}|\;.
\end{align}
For short distances the application of the Mahnattan metric seems reasonable for lattice systems, but for larger distances the system should more and more resemble a continuum, thus there should be a transition from the propagation of the perturbation through the system according to the Manhattan metric to a propagation according to the Euclidian metric also for lattice systems with only nearest-neighbour-interactions. One can only expect to observe this transition in higher-dimensional systems, as for one-dimensional systems there is no difference between the Euclidian and the Manhattan metric and the question concerning the propagation of a perturbation through the system reduces to the determination of its velocity.\\
Former studies on the propagation of a perturbation through a quantum system mainly focused on one-dimensional systems. These systems include the one-dimensional Bose-Hubbard model (BHM) \cite{Krutitsky2014}, the one-dimensional Bose gas \cite{Geiger2014}, the spin-$1/2$ Heisenberg XXZ chain \cite{Bonnes2014} and transverse-field Ising model (TFIM) with long-range interactions \cite{Hauke2013,Cevolani2015} as well as the BHM with long-range interactions \cite{Cevolani2015}. In experiments on trapped ions a finite propagation velocity of a perturbation could also be observed \cite{Cheneau2012,Jurcevic2014}. Considering higher-dimensional systems Cevolani \textit{et al.} in \cite{Cevolani2016} generalized their results for the TFIM and the BHM with long-range interactions in one dimension from \cite{Cevolani2015} to arbitrary lattice dimensions $D$. Their results for the TFIM rely on a spin wave analysis within quadratic approximation, which can be applied deep in the paramagnetic phase. Navez \textit{et al.} have studied the entanglement dynamics of two distant qubits by analyzing correlations in the 2D-TFIM using the large coordination number expansion \cite{Navez2016}. In \cite{Carleo2014} Carleo \textit{et al.} have investigated the spreading of density-density correlations in the BHM on a 1D chain and on a 2D square lattice with only nearest-neighbour-interactions. For the model in two dimensions the geometry of the light cone was studied. Carleo \textit{et al.} have found that the perturbation propagates according to the Manhattan metric, but the studies only considered short distances on the lattice, thus the question of the geometry of the light cone at large distances remains still open for lattice systems with only nearest-neighbour interactions.\\
In this paper we answer this question, considering the 2D-TFIM with only nearest-neighbour-interactions on the square lattice, a well-known many-body standard model of quantum mechanics, for which we study the propagation of a local perturbation. In contrast to the 1D-TFIM, which is integrable and can be solved analytically by a transformation to a system of free fermions \cite{Pfeuty1971}, the 2D-TFIM on the square lattice is non-integrable and cannot be solved exactly. There is \textit{no} transformation to a system of free fermions diagonalizing its Hamiltonian and its relaxation process cannot be described with a semiclassical theory with non-interacting quasiparticles either \cite{Rieger2011,Blass2012}. For this reason we use a time-dependent application of mean-field theory based on the BBGKY (Bogoliubov-Born-Green-Kirkwood-Yvon) hierarchy, which gives an accurate description of the propagation of the perturbation through the lattice and can be applied to the system deep in the ferromagnetic phase. From our numerical data we derive a functional relationship between the propagation velocity of the perturbation and the parameters of the Hamiltonian and study the shape of the light cone, i.e. answer the question whether the Euclidian or the Manhattan metric has to be applied to the 2D-TFIM.

\section{The model}
We study the 2D-TFIM with nearest-neighbour interactions on a square lattice of size $L\times L$ with periodic boundary conditions (PBC), defined by the Hamiltonian
\begin{align}
\hat{H}=-\frac{J}{2}\sum_{<\mathbf{R},\mathbf{R}'>}\hat{\sigma}_{\mathbf{R}}^{x}\hat{\sigma}_{\mathbf{R}'}^{x}-\frac{h}{2}\sum_{\mathbf{R}}\hat{\sigma}_{\mathbf{R}}^{z}\;.
\label{Eq:Hamiltonian}
\end{align}
$J$ is the coupling constant between nearest neighbours and $h$ the transverse field. In the following we set $J=1$ and just vary the transverse field $h$. To describe the state of the system we use the $\mathbf{x}$-basis, in which $\hat{\sigma}_{\mathbf{R}}^{x}$ measures the orientation of the spin at site $\mathbf{R}$ and $\hat{\sigma}_{\mathbf{R}}^{z}$ inverts it. 

\section{Time evolution}
We prepare the system in a generic non-eigenstate of its Hamiltonian and compute the time evolution of the single-particle Bloch vector
\begin{align}
\boldsymbol{S}_{\mathbf{R}}(t)=\braket{\hat{\bm{\sigma}}_{\mathbf{R}}}_{t}
\end{align}
of each spin of the system with a time-dependent application of mean-field theory based on the BBGKY hierarchy \cite{Vardi2001}. The method is very flexible with respect to the initial state of the system and allows us to study large system sizes with more than $10^{4}$ spins (system size $101\times101$). Its accuracy can be controlled by the order of the BBGKY hierarchy.\\
The equations of motion of the Bloch vector of the spin at site $\mathbf{R}$ are derived from the Ehrenfest theorem:
\begin{subequations}
\begin{alignat}{2}
\dot{S}_{\mathbf{R}}^{x}&=&&h\,S_{\mathbf{R}}^{y}\\
\dot{S}_{\mathbf{R}}^{y}&=&&J\,\sum_{k=1}^{2}\big[\braket{\hat{\sigma}_{\mathbf{R}}^{z}\hat{\sigma}_{\mathbf{R}+\mathbf{e}_{k}}^{x}}_{t}+\braket{\hat{\sigma}_{\mathbf{R}}^{z}\hat{\sigma}_{\mathbf{R}-\mathbf{e}_{k}}^{x}}_{t}\big]-h\,S_{\mathbf{R}}^{x}\label{Eq:Ehrenfest_diff_eq_b}\\
\dot{S}_{\mathbf{R}}^{z}&=-&&J\,\sum_{k=1}^{2}\big[\braket{\hat{\sigma}_{\mathbf{R}}^{y}\hat{\sigma}_{\mathbf{R}+\mathbf{e}_{k}}^{x}}_{t}+\braket{\hat{\sigma}_{\mathbf{R}}^{y}\hat{\sigma}_{\mathbf{R}-\mathbf{e}_{k}}^{x}}_{t}\big]
\label{Eq:Ehrenfest_diff_eq_c}
\end{alignat}
\end{subequations}
with $\mathbf{e}_{1}=(1,0)$ and $\mathbf{e}_{2}=(0,1)$. This system of differential equations is not closed, as the two spin correlators are unknown and cannot be computed in a simple way. To obtain a closed system of differential equations, in first order BBGKY hierarchy the general mean-field approximation
\begin{align}
\braket{\hat{\sigma}_{\mathbf{R}}^{y/z}\hat{\sigma}_{\mathbf{R}\pm\mathbf{e}_{k}}^{x}}_{t}\approx\braket{\hat{\sigma}_{\mathbf{R}}^{y/z}}_{t}\braket{\hat{\sigma}_{\mathbf{R}\pm\mathbf{e}_{k}}^{x}}_{t}=S_{\mathbf{R}}^{y/z}\,S_{\mathbf{R}\pm\mathbf{e}_{k}}^{x}
\end{align}
is used, leading to the equations of motion of the bloch vector in first order BBGKY hierarchy:
\begin{subequations}
\begin{alignat}{2}
\dot{S}_{\mathbf{R}}^{x}&=&&h\,S_{\mathbf{R}}^{y}\\
\dot{S}_{\mathbf{R}}^{y}&=&&J\,\sum_{k=1}^{2}\big[S_{\mathbf{R}}^{z}\,S_{\mathbf{R}+\mathbf{e}_{k}}^{x}+S_{\mathbf{R}}^{z}\,S_{\mathbf{R}-\mathbf{e}_{k}}^{x}\big]-h\,S_{\mathbf{R}}^{x}\\
\dot{S}_{\mathbf{R}}^{z}&=-&&J\,\sum_{k=1}^{2}\big[S_{\mathbf{R}}^{y}\,S_{\mathbf{R}+\mathbf{e}_{k}}^{x}+S_{\mathbf{R}}^{y}\,S_{\mathbf{R}-\mathbf{e}_{k}}^{x}\big]\;.
\end{alignat}
\end{subequations}
The equations of motion in second order BBGKY hierarchy can be derived reinserting the two spin correlators in Eqs.\,(\ref{Eq:Ehrenfest_diff_eq_b}) and (\ref{Eq:Ehrenfest_diff_eq_c}) into the Ehrenfest theorem. The expectation values of three Pauli spin operators in the resulting system of nine differential equations are subsequently broken up according to
\begin{align}
\begin{split}
\braket{\hat{A}\hat{B}\hat{C}}_{t}\approx&\braket{\hat{A}\hat{B}}_{t}\braket{\hat{C}}_{t}+\braket{\hat{A}\hat{C}}_{t}\braket{\hat{B}}_{t}+\braket{\hat{B}\hat{C}}_{t}\braket{\hat{A}}_{t}\\
&-2\braket{\hat{A}}_{t}\braket{\hat{B}}_{t}\braket{\hat{C}}_{t}\;.
\end{split}
\end{align}
The expansion to even higher orders follows \cite{Vardi2001,Pucci2016}.\\
\begin{figure}
\begin{center}
\includegraphics{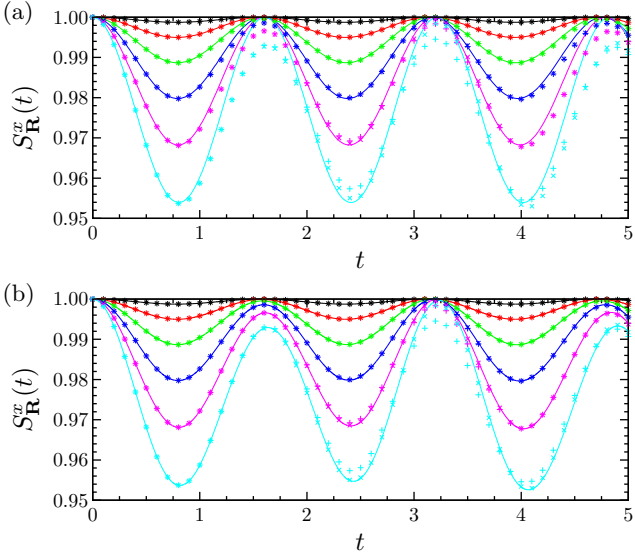}
\end{center}
\caption{Comparison between the results for the $x$-component of the single-particle Bloch vector $S^{x}_{\mathbf{R}}(t)$ of the time-dependent mean-field theory (a) in first and (b) in second order BBGKY hierarchy (continuous lines) to results of rt-VMC ($\boldsymbol{+}$) and time-dependent perturbation theory of 8th order ($\boldsymbol{\times}$). The colour code is as follows: $\textcolor{black}{\boldsymbol{-}}\;h=0.1$, $\textcolor{red}{\boldsymbol{-}}\;h=0.2$, $\textcolor{green}{\boldsymbol{-}}\;h=0.3$, $\textcolor{blue}{\boldsymbol{-}}\;h=0.4$, $\textcolor{mymagenta}{\boldsymbol{-}}\;h=0.5$, $\textcolor{mycyan}{\boldsymbol{-}}\;h=0.6$.}
\label{FIG1}
\end{figure}
\begin{widetext}
\begin{align}
\begin{split}
\scalebox{0.9}{$S^{x}_{\mathbf{R}}(t)=$}&\scalebox{0.9}{$\,1-\left(\tfrac{h}{4J}\right)^{2}\cdot\Big[1-\cos(4Jt)\Big]+\left(\tfrac{h}{4J}\right)^{4}\cdot\Big[\tfrac{16}{9}\cos(6Jt)-\tfrac{7}{3}\cos(4Jt)+\tfrac{16}{3}\cos(2Jt)+6Jt\sin(4Jt)-\tfrac{43}{9}\Big]$}\\
&\hspace*{0.3cm}\scalebox{0.9}{$-\left(\tfrac{h}{4J}\right)^{6}\cdot\Big[\tfrac{13}{32}\cos(8Jt)+\tfrac{34}{3}\cos(6Jt)+\tfrac{137}{18}\cos(4Jt)+\tfrac{26}{9}\cos(2Jt)+\tfrac{5}{4}Jt\sin(8Jt)$}\\
&\hspace*{2cm}\scalebox{0.9}{$+\tfrac{113}{3}Jt\sin(4Jt)+48Jt\sin(2Jt)+24J^{2}t^{2}\cos(4Jt)+3J^{2}t^{2}-\tfrac{2135}{96}\Big]$}\\
&\hspace*{0.3cm}\scalebox{0.9}{$+\left(\tfrac{h}{4J}\right)^{8}\cdot\Big[\tfrac{13}{960}\cos(12Jt)-\tfrac{12287}{8100}\cos(10Jt)+\tfrac{481817}{19200}\cos(8Jt)+\tfrac{401333}{5400}\cos(6Jt)-\tfrac{7469867}{129600}\cos(4Jt)+\tfrac{6062303}{16200}\cos(2Jt)$}\\
&\hspace*{2cm}\scalebox{0.9}{$-\tfrac{196}{135}Jt\sin(10Jt)+\tfrac{6817}{480}Jt\sin(8Jt)+\tfrac{4936}{45}Jt\sin(6Jt)+\tfrac{269717}{4320}Jt\sin(4Jt)+\tfrac{30251}{45}Jt\sin(2Jt)$}\\
&\hspace*{2cm}\scalebox{0.9}{$+\tfrac{281}{48}J^{2}t^{2}\cos(8Jt)-\tfrac{224}{9}J^{2}t^{2}\cos(6Jt)+\tfrac{13213}{72}J^{2}t^{2}\cos(4Jt)-\tfrac{388}{3}J^{2}t^{2}\cos(2Jt)-\tfrac{4801}{144}J^{2}t^{2}$}\\
&\hspace*{2cm}\scalebox{0.9}{$+\tfrac{65}{18}J^{3}t^{3}\sin(8Jt)-3J^{3}t^{3}\sin(4Jt)+\tfrac{1993}{432}J^{3}t^{3}\sin(2Jt)+\tfrac{36}{3}J^{4}t^{4}\cos(4Jt)+\tfrac{4}{3}J^{4}t^{4}-\tfrac{71623969}{172800}\Big]$}\\
&\hspace*{0.3cm}\scalebox{0.9}{$+\mathcal{O}(h^{10})$}
\end{split}
\label{Eq:perturbation_theory}
\end{align}
\end{widetext}
In order to check the accuracy of the approximation we compare the predictions in first and second order BBGKY hierarchy to the corresponding results of real-time Variational Monte Carlo (rt-VMC) as described in \cite{Blass2016} and of time-dependent perturbation theory. For this comparison we consider a spatially homogeneous initial state
\begin{align}
\ket{\psi_{\text{or}}}=\ket{\uparrow\uparrow\uparrow\ldots\uparrow\uparrow}_{x}
\label{ordered_state}
\end{align}
and compute $S^{x}_{\mathbf{R}}(t)$ with the help of rt-VMC and in 8th order perturbation theory given by Eq.\,(\ref{Eq:perturbation_theory}). Note that $S^{x}_{\mathbf{R}}(t)$ is independent of $\mathbf{R}$ as the Hamiltonian as well as the initial state are translation-invariant. Fig.\,\ref{FIG1} shows the comparison of the time evolution of the mean-field prediction to the rt-VMC results and the results of the perturbation theory for $h$ up to $0.6$. For these values of $h$ the system does not leave the ferromagnetic phase as was shown in \cite{Blass2016}. There is a good agreement between the results of the different methods already in first order BBGKY hierarchy, which becomes even better in second order. This agreement is conserved for times much larger than the time interval shown in Fig.\,\ref{FIG1}.\\
Higher orders of BBGKY hierarchy are mandatory to address questions like the nature of the stationary state of the system, while for our considerations of the propagation of the perturbation through the system the first order BBGKY hierarchy would be sufficient, but we will also show results of second order.
\begin{figure*}
\begin{center}
\includegraphics{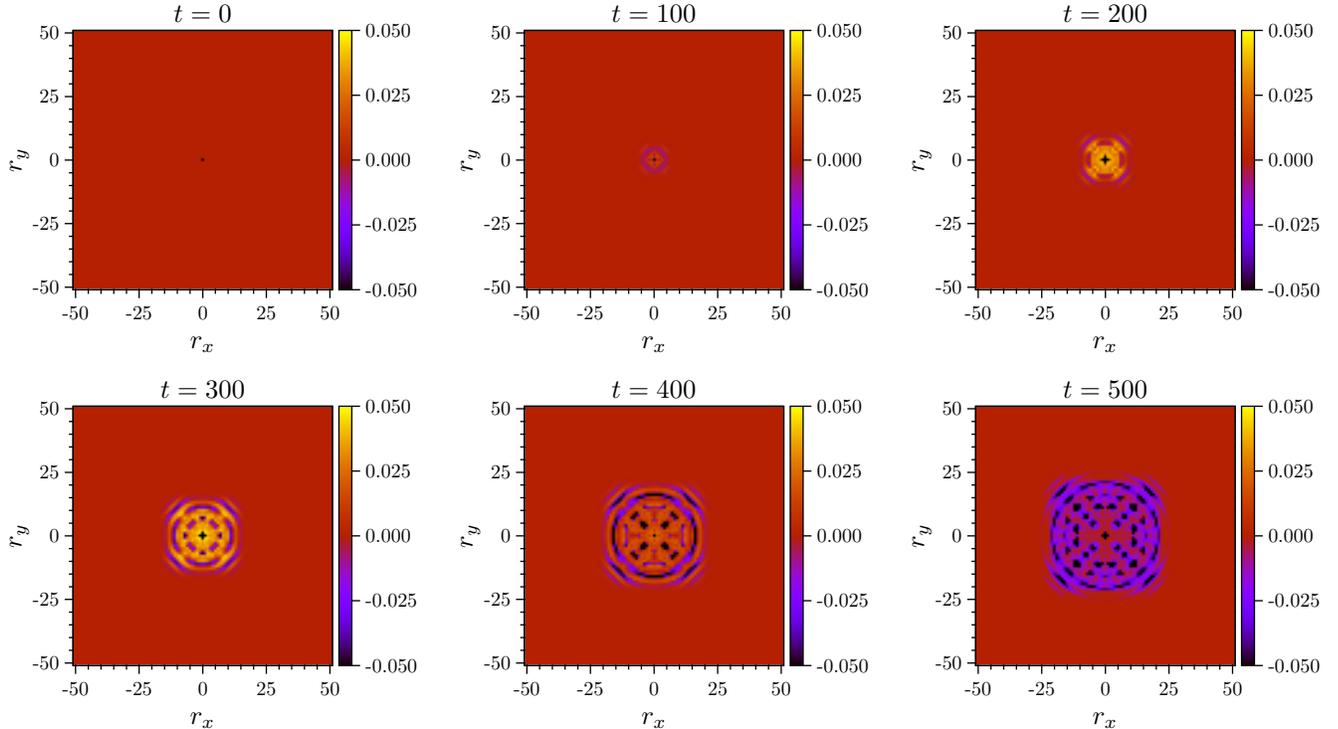}
\end{center}
\caption{$\Delta_{\mathbf{r}}^{x}(t)$ for different times $t$ in the system of size $101\times101$ with $J=1$ and $h=0.6$. At $t=0$ the spin at the site $(0,0)$ is orientated down, while all the other spins of the system are orientated up. For later times the perturbation propagates through the system with a finite velocity according to the Euclidian metric at large distances, defining a light cone with a circular shape.}
\label{FIG2}
\end{figure*}

\section{Results}
\begin{figure}
\begin{center}
\includegraphics{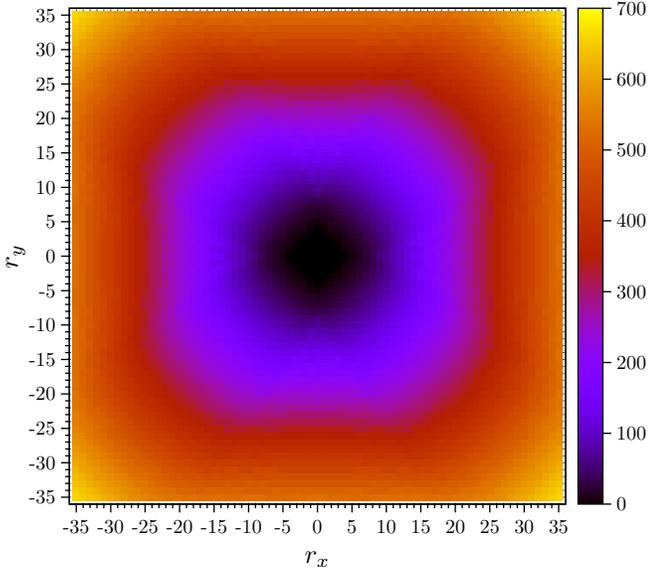}
\end{center}
\caption{Time of arrival of the perturbation for $h=0.6$ in time-dependent mean-field theory of first order BBGKY hierarchy for a subset of the $101\times101$ system. For short distances the propagation of the signal follows the Manhattan metric, while for larger distances the coarseness of the lattice becomes less important and the propagation of the signal is close to the Euclidian metric with an almost circular wave front like in a continuum model.}
\label{FIG3}
\end{figure}

\begin{figure}
\begin{center}
\includegraphics{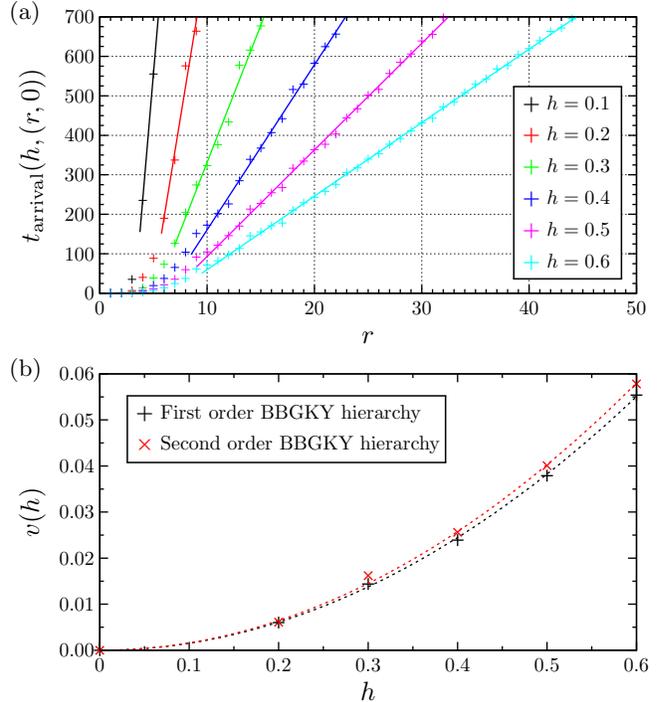}
\end{center}
\caption{(a) Time of arrival of the perturbation along the axis for different values of the transverse field $h$ in time-dependent mean-field theory of first order BBGKY hierarchy in the 2D-TFIM. (b) Velocity of the perturbation along the axis in the 2D-TFIM in time-dependent mean-field theory of first ($\boldsymbol{+}$) and second order BBGKY hierarchy (\textcolor{red}{$\boldsymbol{\times}$}). The propagation velocity increases with $v_{2D}(h)\propto h^{2}$ (dashed lines).}
\label{FIG4}
\end{figure}

We apply a local perturbation in the system by preparing it in a state with one single spin down
\begin{align}
\ket{\psi_{\text{lp}}}=\ket{\downarrow\uparrow\uparrow\ldots\uparrow\uparrow}_{x}\;.
\end{align}
Due to the PBC we may set the initial perturbation at the site $(0,0)$. In order to decide when the perturbation has covered the distance $\mathbf{r}=(r_{x},r_{y})$, we consider the difference
\begin{align}
\bm{\Delta}_{\mathbf{r}}(t)=\braket{\psi_{\text{lp}}|\hat{\bm{\sigma}}_{\mathbf{r}}(t)|\psi_{\text{lp}}}-\braket{\psi_{\text{or}}|\hat{\bm{\sigma}}_{\mathbf{r}}(t)|\psi_{\text{or}}}
\end{align}
to the time evolution starting from the completely ordered state in Eq.\,(\ref{ordered_state}). As the perturbation propagates through the system with a finite velocity, $\bm{\Delta}_{\mathbf{r}}(t)$ vanishes for times when the perturbation has not covered the distance $\mathbf{r}$ yet.\\
We define the time $t_{\text{arrival}}(\mathbf{r})$ when the perturbation has covered the distance $\mathbf{r}$ as the time for which $\bm{\Delta}_{\mathbf{r}}(t)$ becomes non-zero for the first time. Fig.\,\ref{FIG2} shows snapshots of the $x$-component $\Delta_{\mathbf{r}}^{x}(t)$ at different times $t$ for a $101\times101$ system with $h=0.6$. The initial local perturbation propagates through the system with a finite velocity, defining a light cone. The region which has already been passed by the front of the perturbation shows complex amplitude patterns due to interference effects. The light cone geometry of the time-dependent propagation of the initial local perturbation is visualized in Fig.\,\ref{FIG3}, where we show $t_{\text{arrival}}(\mathbf{r})$ for different positions of the system. While for short distances the quadratic shape rotated by $\pi/4$ indicates that the Manhattan metric holds as reported for the Bose Hubbard model in \cite{Carleo2014}, for larger distances the shape of the light cone converges to a circular shape like in Euclidian metric. This means that for large distances the coarseness of the lattice becomes less important and the system behaves like a continuum model as one would expect.\\
The results for $t_{\text{arrival}}$ allow us to determine the propagation velocity of the perturbation via
\begin{align}
v(h)=\frac{d(\mathbf{r})}{t_{\text{arrival}}(h,\mathbf{r})}\;.
\end{align}
Considering the propagation along the axis we can circumvent the question whether the distance $d(\mathbf{r})$ has to be measured in the Euclidian or the Manhattan metric, as for $\mathbf{r}=(r,0)$ we have $d_{\text{Eucl}}(\mathbf{r})=d_{\text{Manh}}(\mathbf{r})=r$. We use the corresponding results for $t_{\text{arrival}}$ to define the propagation velocity $v$ of the perturbation as a function of the transverse field $h$. Fig.\,\ref{FIG4} (a) contains results for $t_{\text{arrival}}$ along the axis as function of the distance $r$ for values of the transverse field up to $h=0.6$ in first order BBGKY hierarchy. $t_{\text{arrival}}$ grows linearly with $r$, i.e. the perturbation propagates uniformly through the lattice. Its velocity is given by the inverse slope of the best fit straight line. In second order BBGKY hierarchy we find similar curves. In Fig.\,\ref{FIG4} (b) we show $v$ as a function of the transverse field $h$ in first and in second order BBGKY hierarchy. For the considered values of the transverse field there are only small differences between the results. In both cases $v$ grows quadratically with $h$, i.e.
\begin{align}
v_{2D}(h)\propto h^{2}\;.
\end{align}
The quadratic dependence of the propagation velocity $v$ of the perturbation on the transverse field $h$ in the 2D-TFIM is in striking contrast to the field-dependence of the propagation velocity in the 1D-TFIM, which is piecewise linear \cite{Rieger2011,Blass2012}:
\begin{align}
v_{1D}(h)=\begin{cases}h&\text{for }h\leq J\\J&\text{for }h>J\end{cases}\;,
\end{align}
i.e. in the ferromagnetic phase the propagation velocity of the perturbation grows linearly with the transverse field $h$ and in the paramagnetic phase it has a constant value given by the coupling constant $J$. Thus for the values of $h$ for which we can determine the propagation of the perturbation in the 2D-TFIM with the time-dependent mean-field theory the propagation velocity for a given value of $h$ is much lower than in the 1D-TFIM. The reason for this is that for the small values of $h$ that we considered the system is deep in the ferromagnetic phase and the coupling between the spins is the dominant contribution to its energy. While in 1D the number of broken bonds in the system, the so-called kinks, is not changed when the spin neighbouring to the initial spin down is flipped, in the 2D-TFIM the described spin flip leads to the creation of two new kinks, which is energetically unfavourable deep in the ferromagnetic phase. In the paramagnetic phase we expect this effect to be less important and thus the difference between the velocity of the propagation of the perturbation between the 1D- and the 2D-TFIM should reduce.

\section{Conclusion and outlook}
Using time-dependent mean-field theory of first and second order BBGKY hierarchy we have shown that a local perturbation in the 2D-TFIM propagates with a finite velocity through the system and obeys the Euclidian metric for large distances from its initial position, resulting in a light cone with a circular shape. This result could have been expected as for large system sizes and large distances the effect of the coarseness of the lattice should vanish and the system should behave like a continuum. On short distances on the other hand we have found that the Manhattan metric holds as one would expect from the lattice structure of the Hamiltonian. For the values of the transverse field which we can simulate the system does not leave the ferromagnetic phase. The propagation velocity of the perturbation increases with $h^{2}$ and is much lower than in the 1D-TFIM, for which the propagation velocity increases linearly with $h$ in the ferromagnetic phase. This can be understood taking into account that in the 2D-TFIM the propagation of the perturbation creates additional kinks which are energetically unfavourable in the ferromagnetic phase. Future studies should aim at increasing the values of the transverse field for which the time evolution of the system can be simulated, thus allowing to study the propagation of the perturbation also in the paramagnetic phase.

\end{document}